\documentstyle[preprint,%
aps]{revtex}
\begin{document}
\draft
\tighten
\preprint{hep-th/9306148; BI-TP 93/25}
\title{Thermal Green's Functions\\
from Quantum Mechanical Path Integrals II:\\
Inclusion of Fermions}
\author{D. G. C. McKeon\cite{email1}}
\address{Department of Applied Mathematics, University of Western Ontario,\\
London, Ontario, Canada N6A 5B7}
\author{A. K. Rebhan\cite{byline}}
\address{Fakult\"at f\"ur Physik, Universit\"at Bielefeld,\\
D--33501 Bielefeld, Germany
}
\date{June 23, 1993}
\maketitle
\begin{abstract}
In a previous paper we have shown how, for bosonic fields,
the generating functional in both relativistic quantum field
theory and thermal field theory can be evaluated by use of a
standard quantum mechanical path integral. In this paper we extend
this method to include fermionic fields. A particular problem
is posed by Green's functions
with external fermionic lines,
where the different boundary conditions
of bosons and fermions in imaginary time
have to be accommodated within one
path integral expression. The general procedure
is worked out in the example of
scalar and spinor self-energies in a simple model
with a Yukawa coupling of a scalar to a Majorana spinor.
\end{abstract}

\narrowtext
\section{Introduction}

Loop corrections in thermal field theory can be calculated
using a Feynman diagrammatic approach which is similar to the
way in which analogous processes can be computed in ordinary
quantum field theory \cite{Kapusta}. In the Matsubara formulation
of thermal field theory, the space is Euclidean rather than
Minkowski with (anti-)periodic boundary conditions in imaginary
time with extent $\beta=1/(kT)$, where $T$
is the temperature. Thus the frequencies of all fields
are either integer multiples of $2\pi/\beta$ (for Bose particles)
or half-integer multiples (for Fermi particles). Apart from
having sums over frequencies instead of integrals, the perturbation
theory is the same as at zero temperature.

Recently an alternative to the Feynman diagrammatic approach
has been proposed \cite{G} in the context of operator regularization
\cite{OR}. In the latter, the evaluation of Green's functions
is based on the computation of a matrix element of the form
\begin{equation}
M_{xy}= \langle x|e^{-iHt}|y\rangle
\label{M}\end{equation}
in powers of the background fields upon which the ``Hamiltonian'' $H$
depends. If this is done by using the ``Schwinger expansion''
\cite{Schwinger}
\begin{eqnarray}
e^{A+B}=\sum_{n=1}^\infty
\int_0^\infty d\alpha_1\cdots d\alpha_n
&&\,\delta(1-\alpha_1-\alpha_2\ldots-\alpha_n)\nonumber\\
\times&&e^{\alpha_1 A}Be^{\alpha_2 A}\cdots Be^{\alpha_n A}
\label{Schwexp}\end{eqnarray}
then the calculation is quite similar in form to what is encountered
in the conventional Feynman diagram approach; both loop-momentum
and Feynman parameter integrals have to be evaluated.
The alternative approach based on quantum mechanical path integrals
\cite{G} computes Eq.~(\ref{M}) more directly, and turns out to
obviate loop-momentum integrals completely.

This latter method has been applied to calculations in relativistic
quantum field theoretical models involving scalars, spinors, and
gauge fields to one and two loop order \cite{G}.\footnote{A similar
approach has been developed independently, at one loop order,
in Ref.\ \cite{Stra}.} In a previous paper
we have shown how this approach can be used to evaluate Green's
functions in thermal field theory in the purely bosonic case \cite{I}.
Again, no loop integrals over spatial momenta are encountered,
whereas in place of the sum over Matsubara frequencies one has
to perform a sum over winding numbers of paths with respect to
the circular imaginary time, the two conceptually different
sums being related by Jacobi's
imaginary transformation for theta functions. The representation
as a sum over winding numbers was shown to also constitute an
interesting alternative for the derivation of high-temperature
expansions of Green's functions.

In this paper we consider the generalization of this alternative
approach to include
fermions, which differ in their boundary conditions in imaginary time,
being antiperiodic rather than periodic.
In the case that only fermions are involved, it is not surprising
to find that
the only difference consists of having to evaluate an alternating
sum over winding numbers of paths in imaginary time,
which, through the Jacobi transformation, corresponds
to the usual sum over half-integer
Matsubara frequencies. However,
is not so obvious how the different boundary conditions have to
be accounted for if both fermionic and bosonic fields appear in
a single loop diagram, which is to be replaced by a certain
quantum mechanical path integral.
To illustrate how this is done, we calculate the two-point functions
at one-loop order in a simple model in which a scalar field is coupled to
a Majorana spinor by a Yukawa interaction. The generalization to
more complicated (and more interesting) field theories is
entirely straightforward, given the methods described in Ref.\
\cite{G}.

\section{Thermal Green's Functions Involving Fermions}

We consider a four-dimensional model for a scalar field $A$ and a
Majorana spinor $\chi$ with Lagrangian density
\begin{eqnarray}
{\cal L}&=&-\frac12 Ap^2A-\frac i2 \bar\chi p\llap/ \chi
-\frac12 \lambda \bar\chi A\chi \label{L}\\ \nonumber
&&(p=-i\partial, \bar\chi=\chi^{\rm T}C ).
\end{eqnarray}
In the imaginary-time formulation of thermal field theory \cite{Kapusta},
the four-dimensional space is Euclidean, $x^0$ having the finite
range $0\le x^0\le\beta$, with Bose (Fermi) fields subject to
periodic (anti-periodic) boundary conditions.

We shall consider scalar and fermionic Green's functions in turn,
which --- in the diction of the background field method \cite{BFM} ---
means that we have to introduce background fields for $A$ and $\chi$,
respectively, while the other field appears only as a quantum field.

\subsection{Scalar Self-Energy}

With the spinor field $\chi$ being a purely quantum field and
$A$ having a background contribution $B$, the generating functional
for scalar Green's functions at one-loop order reads
\begin{eqnarray}
Z^{(1)}[B]
&=&{\rm Sdet}^{-\frac12}
\left(\begin{array}{cc}
-p^2 & 0\\
0 & -(ip\llap/+\lambda B) \end{array}\right)\nonumber\\
&=&{\rm const.}\times{\rm Det}^{\frac12}(ip\llap/+\lambda B),
\label{Zb}\end{eqnarray}
where ``Sdet'' denotes the functional superdeterminant \cite{DeW} with
the appropriate boundary conditions, which, for the final,
ordinary functional determinant
in Eq.~(\ref{Zb}) are antiperiodic ones, because the latter
is the fermionic part of the former.

Introducing the matrix operator
\begin{eqnarray}
H&=&-(ip\llap/+\lambda B)^2
=p^2-i\lambda(p\llap/ B+B p\llap/)-\lambda^2 B^2\nonumber\\
&=&(p-i\lambda\gamma B)^2-3\lambda^2 B^2,
\label{Hb}\end{eqnarray}
we regulate Eq.~\ref{Zb} by the $\zeta$ function \cite{OR,zeta}
\begin{equation}
\zeta(s)={\rm Tr}\frac1{\Gamma(s)}
\int_0^\infty dt\,t^{s-1}e^{-Ht}
\label{zetaHb}\end{equation}
according to $\ln Z^{(1)}[B]=\frac14\zeta'(0)$.
We therefore have to evaluate
\begin{equation}
{\rm Tr}\,\exp(-Ht)=\int d^4x\,d^4y\,\delta(x-y)\,
\langle x|{\rm tr}\,\exp(-Ht)|y\rangle.
\label{trexp}\end{equation}
While at one-loop order only the diagonal matrix elements enter,
at higher loop orders the general matrix elements are the basic building
blocks of the generating functional, if defined
by the method of operator regularization \cite{OR}.

In order to obtain the one-loop contribution to the
(thermal) scalar self-energy,
only the term bilinear in the background field $B$
needs to be extracted from
Eq.~(\ref{trexp}).

\subsubsection{Evaluation by Schwinger expansion}

We first evaluate Eq.~(\ref{trexp}) by employing the
Schwinger expansion, Eq.~(\ref{Schwexp}), which reproduces
the expressions derived from the more customary Feynman rules
together with Feynman parametrization of denominators. This
will serve as a reference for the results to be derived from
the use of quantum mechanical path integrals.

To second order in $B$, Eq.~(\ref{Schwexp}) yields
\widetext
\begin{equation}
{\rm Tr}\,\exp(-Ht)\bigg|_{B^2}=\lambda^2
{\rm Tr}\biggl[ t B^2 e^{-p^2t}-\frac{t^2}2
\int_0^1 d\alpha\,e^{-(1-\alpha)p^2t}(p\llap/ B+B p\llap/)
e^{-\alpha p^2t}(p\llap/ B+B p\llap/)\biggr].
\label{trexp2}\end{equation}

Evaluating further by insertion momentum states $|p\rangle$,
we have, remembering the antiperiodic boundary conditions for the
functional determinant in Eq.~(\ref{Zb}) and hence for the
functional trace in the subsequent expressions,
\begin{mathletters}
\begin{equation}
\langle x|p\rangle =\frac1{(2\pi)^{3/2}\sqrt{\beta}}
\exp i \left[ {\bf px}+\frac{2\pi}{\beta}(n_p+\frac12)x_4\right]
\end{equation}
and
\begin{equation}
\langle p|B|q\rangle =\frac1{(2\pi)^{3/2}\sqrt{\beta}}
B({\bf p}-{\bf q},\frac{2\pi}{\beta}(n_p-n_q)),
\end{equation}
\end{mathletters}
with $n_p$ and $n_q$ integers. After appropriate shifts of the
integration and summation variables, and then using
$\int e^{-{\bf q}^2t}d^3q/(2\pi)^3=(4\pi t)^{-3/2}$, we end up with
\begin{eqnarray}
&&{\rm Tr}\,\exp(-Ht)\bigg|_{B^2}=-4\lambda^2
\int\frac{d^3p}{(4\pi t)^{3/2}\beta}\sum_{n_pn_q}\biggl\{
-te^{-(\frac{2\pi}{\beta})^2(n_q+\frac12)^2t} \nonumber\\
&&\qquad\qquad\qquad\qquad\qquad\qquad
+t^2\int_0^1 d\alpha\,e^{-\alpha(1-\alpha){\bf p}^2t
-(\frac{2\pi}{\beta})^2[\alpha(n_p+n_q+\frac12)^2+
(1-\alpha)(n_q+\frac12)^2]t} \biggr\} \nonumber\\
&&\times B({\bf p},\frac{2\pi}{\beta}n_p)
\left[\frac3t-\alpha(1-2\alpha){\bf p}^2+
(\frac{2\pi}{\beta})^2(n_p(n_q+\frac12)+2(n_q+\frac12)^2)\right]
B(-{\bf p},-\frac{2\pi}{\beta}n_p).
\label{trexp2r}\end{eqnarray}
In accordance with the Feynman rules in the Matsubara formalism,
we encounter imaginary frequencies for the scalar background fields
which are integer multiples of $2\pi/\beta$, and there is a
further summation over half-integer ($n_q+\frac12$) multiples
corresponding to the frequencies of the internal fermion lines.

\subsubsection{Evaluation by a quantum mechanical path integral}

We now compute the matrix element appearing in Eq.~(\ref{trexp})
by using the representation through the standard quantum mechanical
path integral \cite{Feynman}
\begin{eqnarray}
&&\langle x| \exp\left\{ -[\frac12(p-{\cal A})^2+{\cal V}]t\right\}|y\rangle
\nonumber\\
&=&{\rm I\!P} \int Dq \exp\int_0^t d\tau
\left[ -\frac12 \dot q^2(\tau)+i\dot q(\tau)\cdot{\cal A}(q(\tau))
-{\cal V}(q(\tau) \right],
\label{pint}\end{eqnarray}
with $q(0)=y$ and $q(t)=x$. ${\rm I\!P}$ stands for path ordering in the
case that $\cal A$ and $\cal V$ are matrix valued.

Comparing with Eq.~(\ref{Hb}) and rescaling $t\to t/2$ for convenience,
we identify ${\cal A}=i\lambda\gamma B$ and ${\cal V}=\frac32\lambda^2 B^2$.
Consequently we have
\begin{equation}
{\rm Tr}\,\exp(-Ht)
={\rm Tr}\,{\rm I\!P} \int Dq \exp\int_0^t d\tau
\left[ -\frac12 \dot q^2(\tau) -\dot q(\tau)\cdot\gamma B
-\frac32\lambda^2 B^2 \right].
\end{equation}
To second order in $B$ this becomes
\begin{eqnarray}
{\rm Tr}\,\exp(-Ht)\bigg|_{B^2}&=&\lambda^2{\rm Tr}\int Dq\, e^{-\frac12
\int_0^t d\tau\dot q^2(\tau)}
\biggl\{ -\frac32\int_0^t d\tau_1\,B^2(q(\tau_1)) \nonumber\\
&&+\int_0^t d\tau_1 \int_0^{\tau_1}d\tau_2\,
\dot q(\tau_1)\cdot\gamma B(q(\tau_1))\,
\dot q(\tau_2)\cdot\gamma B(q(\tau_2)) \biggr\},
\label{pb2}\end{eqnarray}
where in the last term
the integration over $\tau_2$ is restricted by the necessity
of path ordering of the $\gamma$ matrices.\footnote{An
alternative to explicit path ordering of $\gamma$ matrices, which
moreover removes their explicit appearance altogether, could be
given by the introduction of fermionic path integration variables
in addition to $q$ \cite{GAMMA}. However, this has no bearing on
the problem of boundary conditions in imaginary time, which is our
main concern here.}

If we now take $B$ to be a plane wave field,
\begin{equation}
B(q(\tau_i))=\frac{b_i}{(2\pi)^{3/2}\sqrt{\beta}}
e^{ik_i\cdot q(\tau_i)}
\equiv\frac{b_i}{(2\pi)^{3/2}\sqrt{\beta}}
e^{i\left[
{\bf k}_i{\bf q}(\tau_i)+(\frac{2\pi}{\beta})n_i q_4(\tau_i)\right]},
\end{equation}
and assign labels 1 and 2 to the $\gamma$ matrices contracted with
$\dot q(\tau_1)$ and $\dot q(\tau_2)$, respectively,
then we can write
\begin{eqnarray}
{\rm Tr}\,\exp(-Ht)\bigg|_{B^2}&=&
\frac{\lambda^2 b_1b_2}{(2\pi)^3\beta}{\rm Tr}\int Dq\, e^{-\frac12
\int_0^t d\tau\dot q^2(\tau)}
\biggl\{ -\frac32\int_0^t d\tau_1\, e^{\int_0^t d\tau
\left[i q(\tau)\cdot (k_1+k_2)\delta(\tau-\tau_1)\right]} \nonumber\\
&&+\int_0^t d\tau_1 \int_0^{\tau_1}d\tau_2\, e^{ \int_0^t d\tau
\left[ {q}(\tau)\cdot\left( i{k}_1 \delta(\tau-\tau_1)+
i{k}_2 \delta(\tau-\tau_2)-\gamma_1\delta'(\tau-\tau_1)
-\gamma_2\delta'(\tau-\tau_2)\right)  \right]},
\end{eqnarray}
provided we keep only the linear terms in $\gamma_1$
and $\gamma_2$.

\narrowtext
The path integral is now of the form
\begin{equation}
P^\beta_{(x;y)}[\Gamma]=
\int Dq \exp \int_0^t d\tau \left[-\frac12\dot q(\tau)+
q(\tau)\cdot \Gamma(\tau) \right],
\label{PG}\end{equation}
where the superscript $\beta$ reminds of the fact that
$q_4$ lives on a circle with circumference $\beta$, i.e.\
$q_4\equiv q_4+n\beta$, with $n$ integer.
This topological constraint prevents us from a direct evaluation
by a shift of variable
\begin{equation}
\tilde q(\tau)=q(\tau)-\int_0^t d\tau' G(\tau,\tau')\Gamma(\tau')
\label{qshift}\end{equation}
as in zero-temperature field theory \cite{G}. In the latter,
\begin{equation}
G(\tau,\tau')=\frac12|\tau-\tau'|-\frac12(\tau+\tau')+\frac{\tau\tau'}t
\label{Gtt}\end{equation}
would render the path integral purely Gaussian.

The problem is that of a path integral quantization of a particle on
a circle, and the general solution has been given in Ref.~\cite{Schulman}
in terms of a superposition of unconstrained path integrals,
\begin{equation}
P^\beta_{(x;y)}[\Gamma]=\sum_{n=-\infty}^{\infty}e^{i\delta_n}
P^\infty_{({\bf x},x_4+n\beta;y)},
\label{Psum}\end{equation}
which corresponds to summing over all paths from $y$ to $x$ which
have different winding number in the fourth variable.

In the purely bosonic case treated in Ref.~\cite{I}, the phases
$\delta_n$ could be set to zero. The present path integral, however,
is supposed to give the matrix elements for taking a functional
trace with antiperiodic boundary conditions in imaginary time.
We therefore have to require that $P^\beta_{(x;y)}$ picks up
a phase equal to $-1$ when making one turn in the circular
imaginary time. This determines $\delta_n=n\pi$, and $P^\beta$
is given by an alternating sum over winding numbers.

\widetext
On the left hand side of Eq.~(\ref{Psum}), we can indeed
perform the shift of Eq.~(\ref{qshift}), yielding
\begin{eqnarray}
P^\infty_{(x;y)}[\Gamma]&=&\frac1{(2\pi t)^2}\exp\biggl[
-\frac{(x-y)^2}{2t}+\frac1t \int_0^t d\tau[x\tau
+y(t-\tau)]\cdot \Gamma(\tau)\nonumber\\
&&\qquad-\frac12\int_0^t d\tau d\tau' G(\tau,\tau')
\Gamma(\tau)\cdot\Gamma(\tau')\biggr],
\label{Pinfty}\end{eqnarray}
with $G$ given by Eq.~(\ref{Gtt}).

Evaluating the above path integral in this manner and computing
the trace by integrating over $x$ after setting $x=y$, which has
the effect of equating $k_1=-k_2\equiv k$, we obtain
\begin{eqnarray}
&&{\rm Tr}\,\exp(-Ht)\bigg|_{B^2}=
\frac{2\lambda^2 b_1b_2}{\beta(2\pi t)^2}\sum_n (-1)^n e^{-n^2\beta^2/(2t)}
\biggl\{ t \nonumber\\
&&+\int_0^t d\tau_1 \int_0^{\tau_1} d\tau_2\,
e^{-k^2\alpha(1-\alpha)t-in\beta k_4\alpha}
\left[ -\frac4t-{\bf k}^2(\alpha-\frac12)^2
+(\frac{n\beta}t-ik_4(\alpha-\frac12))^2\right]\biggr\},
\label{pisres}\end{eqnarray}
where $\alpha\equiv(\tau_1-\tau_2)/t$. Because $k_4=
\frac{2\pi}{\beta}n_k$ with integer $n_k$, we find that
the integrand $f(\alpha)$ in Eq.~(\ref{pisres}) is symmetric under
$\alpha \leftrightarrow 1-\alpha$ together with
$n \leftrightarrow -n$, which allows us to
simplify
\begin{equation}
\int_0^t d\tau_1 \int_0^{\tau_1} d\tau_2 f(\alpha)
\to t^2\int_0^1 d\alpha (1-\alpha) f(\alpha) =
\frac{t^2}2 \int_0^1 d\alpha f(\alpha).
\end{equation}

\narrowtext
Eq.~(\ref{pisres}) involves an infinite sum different from the one
encountered in the result of the more conventional Schwinger expansion,
Eq.~(\ref{trexp2r}).
Introducing Jacobi's $\theta$ functions \cite{HTF}
\begin{mathletters}\label{th234}
\begin{eqnarray}
\theta_2(z|\tau)&=&\sum_{n=-\infty}^\infty \exp\Bigl(
i\pi[\tau (n+\frac12)^2+2z(n+\frac12)]\Bigr), \\
\theta_3(z|\tau)&=&\sum_{n=-\infty}^\infty \exp\left(
i\pi[\tau n^2+2zn]\right), \\
\theta_4(z|\tau)&=&\sum_{n=-\infty}^\infty (-1)^n\exp\left(
i\pi[\tau n^2+2zn]\right),
\end{eqnarray}
the path integral result Eq.~(\ref{pisres}) can be expressed in
terms of $\theta_4$ and its derivatives with respect to its two
arguments, whereas the conventional result Eq.~(\ref{trexp2r})
requires $\theta_2$. Jacobi's so-called imaginary transformation
\end{mathletters}
\begin{equation}
\theta_{3+i}(z|\tau)=(-i\tau)^{-\frac12}e^{-i\pi z^2/\tau}
\theta_{3-i}\Bigl(\frac{z}{\tau}\Big|-\frac1{\tau}\Bigr),
\label{Jit}\end{equation}
where $i=-1,0,$ or $1$, relates $\theta_4$ to $\theta_2$,
and indeed, after some algebra,
establishes the equivalence of the two results.

\subsection{Spinor Self-Energy}

We shall now consider the case of Green's functions
with external fermionic lines, which receive radiative corrections
from both (quantum) scalar and fermionic degrees of freedom
at the same time.
By taking now $A$ to be a purely quantum field and $\chi$
having a background part $\eta$,
the generating functional for fermionic Green's functions
at one-loop order reads
\begin{equation}
Z^{(1)}[\eta]={\rm Sdet}^{-\frac12}{\sf Y}[\eta]
={\rm Sdet}^{-\frac12}
\left(\begin{array}{cc}
p^2 & \lambda\bar\eta \\
\lambda\eta & ip\llap/ \end{array}\right).
\label{Zf}\end{equation}

Dropping an irrelevant constant over-all factor, we may rescale
the supermatrix ${\sf Y}$ in Eq.~(\ref{Zf}) with the
field-independent supermatrix
\begin{equation}
{\sf X}=\left(\begin{array}{cc}
1 & 0 \\
0 & -ip\llap/ \end{array}\right),
\label{X}\end{equation}
and regulate the superdeterminant of ${\sf H}={\sf X Y}$
using the $\zeta$ function
\begin{equation}
\zeta(s)={\rm Str}\frac1{\Gamma(s)}
\int_0^\infty dt\,t^{s-1}e^{-{\sf H}t}
\label{zetaH}\end{equation}
according to $\ln Z^{(1)}[\eta]=-\frac12\zeta'(0)$.
We thus are confronted with having to evaluate
\widetext
\begin{equation}
{\rm Str}\exp(-{\sf H}t)
=\int d^4x\,d^4y\,\delta(x-y)\,
\langle x|{\rm str}\,\exp\left[-
\left(\begin{array}{cc}
p^2 & \lambda\bar\eta \\
-i\lambda p\llap/\eta & p^2 \end{array}\right)t\right]|y\rangle,
\label{strexp}\end{equation}
where ``Str'' denotes the supertrace including
a functional trace with the appropriate boundary conditions
and ``str'' the
supertrace in the purely algebraic sense.

In order to obtain the one-loop contribution to the
(thermal) spinor self-energy,
we now have to extract the term bilinear in $\eta$ from
Eq.~(\ref{strexp}).

\subsubsection{Evaluation by Schwinger expansion}

Again, we first lay out the derivation of
the thermal spinor self-energy
employing the Schwinger expansion, Eq.~(\ref{Schwexp}),
by which we formally have
\begin{equation}
\langle x|{\rm str}\exp(-{\sf H}t)|y\rangle\bigg|_{\bar\eta\eta}=
-i\lambda^2t^2\int_0^1d\alpha
\langle x|e^{-(1-\alpha)p^2t}\bar\eta p\llap/ e^{-\alpha
p^2t} \eta|y\rangle.
\end{equation}
In order to correctly
take into account
the (anti-)periodic boundary conditions of the supermatrix operator
$\sf H$ when computing the functional trace in
Eq.~(\ref{strexp}) in momentum-space,
one has to remember that $\bar\eta$ appears in its
Bose-Fermi block, and $\eta$ in the Fermi-Bose one.
Therefore,
\begin{equation}
\langle p|\bar\eta|q\rangle=\frac1{\sqrt{(2\pi)^2\beta}}
\bar\eta\bigl({\bf p}-{\bf q},\frac{2\pi}{\beta} n_p-
\frac{2\pi}{\beta} (n_q+\frac12)\bigr),
\label{petaq}\end{equation}
with $n_p$ and $n_q$ integers,
which leaves us with
\begin{eqnarray}
{\rm Str}\exp(-{\sf H}t)\bigg|_{\bar\eta\eta}=&&
-i\lambda^2t^2\int_0^1d\alpha\int\frac{d^3p\,d^3q}{(2\pi)^3\beta}
\sum_{n_pn_q}e^{-[(1-\alpha)({\bf p}^2+(2\pi n_p/\beta)^2
+\alpha({\bf q}^2+(2\pi(n_q+\frac12)/\beta)^2)]t}\nonumber\\
\times\bar\eta\bigl({\bf p}-{\bf q},\frac{2\pi}{\beta}&&(n_p-n_q-\frac12)
\bigr)\left[ {\bf q}\llap/ + \frac{2\pi}{\beta}(n_q+\frac12)\gamma_4
\right] \eta\bigl({\bf q}-{\bf p},\frac{2\pi}{\beta}(n_q+\frac12-n_p)
\bigr).
\label{strexp2}
\end{eqnarray}
With the shifts ${\bf p}\to{\bf p}+{\bf q}$, $n_p\to n_p+n_q$, followed
by ${\bf q}\to{\bf q}-\alpha{\bf p}$, and then using the integral
$\int e^{-{\bf q}^2t}d^3q/(2\pi)^3=(4\pi t)^{-3/2}$, Eq.~(\ref{strexp2})
reduces to
\begin{eqnarray}
{\rm Str}\exp(-{\sf H}t)\bigg|_{\bar\eta\eta}=&&
-i\lambda^2t^2\int_0^1d\alpha\int\frac{d^3p}{(4\pi t)^{3/2}\beta}
\sum_{n_pn_q}e^{-[\alpha(1-\alpha){\bf p}^2+(2\pi/\beta)^2
\{(1-\alpha)(n_p+n_q)^2+\alpha(n_q+\frac12)^2\}]t}\nonumber\\
\times\bar\eta\bigl({\bf p},\frac{2\pi}{\beta}&&(n_p-\frac12)
\bigr)\left[ -(1-\alpha)
{\bf p}\llap/ + \frac{2\pi}{\beta}(n_q+\frac12)\gamma_4
\right] \eta\bigl(-{\bf p},-\frac{2\pi}{\beta}(n_p-\frac12)\bigr).
\label{fres}
\end{eqnarray}

\subsubsection{Evaluation by a quantum mechanical path integral}

In order to represent Eq.~(\ref{strexp}) by the standard
quantum mechanical path integral Eq.~(\ref{pint}), we rewrite
\begin{equation}
{\sf H}=
\left(\begin{array}{cc}
p^2 & \lambda\bar\eta \\
-i\lambda p\llap/\eta & p^2 \end{array}\right)=
\left(\begin{array}{cc}
p_\mu & 0 \\
-\frac{i\lambda}2 \gamma_\mu \eta & p_\mu \end{array}\right)^2+
\left(\begin{array}{cc}
0 & \lambda\bar\eta \\
-\frac{\lambda}2 \gamma_\nu \eta_{,\nu} & 0 \end{array}\right),
\end{equation}
leading to
\begin{eqnarray}
&&\langle x|\exp(-{\sf H}t/2)|y\rangle \nonumber\\
&=&{\rm I\!P}\int Dq\exp\int_0^t d\tau\left[
-\frac12\dot q^2+i\dot q_\nu
\left(\begin{array}{cc}
0 & 0 \\
\frac{i\lambda}2 \gamma_\nu \eta & 0 \end{array}\right)-
\left(\begin{array}{cc}
0 & \frac{\lambda}2 \bar\eta \\
-\frac{\lambda}4 \gamma_\nu \eta_{,\nu} & 0 \end{array}\right)\right],
\label{fpi}\end{eqnarray}
where we have again rescaled $t\to t/2$ in Eq.~(\ref{strexp})
in order to employ the standard quantum mechanical result
Eq.~(\ref{pint}) directly.

To second order in $\eta$, $\bar\eta$ we find from Eq.~(\ref{fpi})
that
\begin{eqnarray}
&&\langle x|{\rm str}\exp(-{\sf H}t/2)|y\rangle\bigg|_{\bar\eta\eta}
\nonumber\\
&=&\int Dq\, e^{-\frac12\int_0^t d\tau\dot q^2(\tau)}
{\rm str}\left[\frac1{2!}{\rm I\!P}\prod_{i=1}^2 \int_0^t d\tau_i
\left(\begin{array}{cc}
0 & -\frac{\lambda}2 \bar\eta_i \\
-\frac{\lambda}2 \dot {q\llap/}_i \eta_i
+\frac{\lambda}4 \gamma_\nu \eta_{i,\nu} & 0 \end{array}\right)\right],
\end{eqnarray}
where $q_i\equiv q(\tau_i)$ and $\eta_i\equiv\eta(q(\tau_i))$.

Computing the (algebraic) supertrace gives two contributions,
the trace over the bosonic part minus the one over the fermionic part,
which are almost identical apart from the sign,
were it not for the path ordering
prescription ${\rm I\!P}$, which is
required because we are dealing with a matrix valued
``Hamiltonian'' ${\sf H}$. The bosonic part of the supertrace reads
\begin{eqnarray}
&&\langle x|{\rm tr_B}\exp(-{\sf H}t)|y\rangle\bigg|_{\bar\eta\eta}
\nonumber\\
&=&\int Dq\, e^{-\frac12\int_0^t d\tau\dot q^2(\tau)}
\frac{\lambda^2}4 \int_0^t d\tau_1\int_0^{\tau_1}d\tau_2
\bar\eta_1\left( \dot {q\llap/}_2 \eta_2
-\frac12 \gamma_\nu \eta_{2,\nu} \right),
\label{trB}\end{eqnarray}
whereas the fermionic part equals minus this expression where
the indices 1 and 2 are interchanged except in the integration
symbols.

Without the topological constraint of a circular imaginary time $q_4$,
the bosonic and the fermionic trace could be simply combined,
resulting in a symmetrized integration domain in $\tau_1$ and
$\tau_2$. However, in the preceding section we have seen that
the path integral corresponding to a fermionic trace when
expressed in terms of an unconstrained path integral, acquires
a nontrivial phase for odd winding numbers in imaginary time,
whereas the bosonic part does not. The two pieces have therefore to
be treated as two different path integral expressions in the
thermal case.

Keeping this difference in mind, we can transform Eq.~(\ref{trB})
and its fermionic counterpart into a Gaussian path integral
by taking the background wave functions to be plane wave fields
whose frequencies are half integral multiples of $2\pi/\beta$,
\begin{equation}
\eta(q(\tau_i))=\frac{u_i}{(2\pi)^{3/2}\sqrt{\beta}}
e^{ik_i\cdot q(\tau_i)}
\equiv\frac{u_i}{(2\pi)^{3/2}\sqrt{\beta}}
e^{i\left[
{\bf k}_i{\bf q}(\tau_i)+(\frac{2\pi}{\beta})(n_i
+\frac12) q_4(\tau_i)\right]}.
\end{equation}
This reduces Eq.~(\ref{trB}) to
\begin{eqnarray}
&&\langle x|{\rm tr_B}\exp(-{\sf H}t)|y\rangle\bigg|_{\bar\eta\eta}
\nonumber\\
&=&\frac{\lambda^2}{4(2\pi)^3\beta}\int_0^t d\tau_1\int_0^{\tau_1}
d\tau_2 \bar u_1 \int Dq \biggl\{ e^{\int_0^t d\tau \left[
-\frac12 \dot q^2(\tau)+q(\tau)\cdot
\left( -i k_1 \delta(\tau-\tau_1)+ik_2\delta(\tau-\tau_2)-
\gamma \delta'(\tau-\tau_2) \right)\right]} \nonumber\\
&&\qquad\qquad
-\frac{i}2 {k\llap/}_2
e^{\int_0^t d\tau \left[
-\frac12 \dot q^2(\tau)+q(\tau)\cdot
\left( -i k_1 \delta(\tau-\tau_1)+ik_2\delta(\tau-\tau_2) \right)\right]}
\biggr\}_{\rm lin.}u_2,
\label{trBu}\end{eqnarray}
where the subscript ``lin.'' indicates that we have exponentiated
$\gamma$ matrices with the provision that
only first order terms in $\gamma$ are to be kept.

Eq.~(\ref{trBu}) is now of the form of Eq.~(\ref{PG}) with
linear sources
$\Gamma=-i k_1 \delta(\tau-\tau_1)+ik_2\delta(\tau-\tau_2)$
in the first term, and
$\Gamma'=\Gamma-\gamma \delta'(\tau-\tau_2)$ in the second one.
The fermionic part of the supertrace gives rise to a similar
expression with source terms in which the labels 1 and 2 are
interchanged.
The two path integral contributions can now be evaluated by
Eq.~(\ref{Pinfty}) and Eq.~(\ref{Psum}), taking into account
the alternating signs picked up by the one corresponding to
the fermionic part of the supertrace.
Performing also the integration over $x$ after setting $x=y$,
which enforces momentum conservation so that $k_1=k_2=k$,
we arrive at the result
\begin{eqnarray}
{\rm Str}\exp(-{\sf H}t/2)&&=
\frac{\lambda^2}{4(2\pi t)^2}\sum_n
\int_0^1 d\sigma_1 \int_0^{\sigma_1}d\sigma_2
e^{-\frac{n^2\beta^2}{2t}-\frac12 k^2
( \alpha-\alpha^2 )t }\nonumber\\
\bar u(-k)\biggl\{&&
e^{-in\beta k_4 \alpha}\left[
e^{n\beta\gamma_4/t+\frac{i}2 k\llap/ (2\alpha-1)}
-\frac{i}2 k\llap/ \right]\nonumber\\
+(-1)^n&& e^{in\beta k_4 \alpha}\left[
e^{n\beta\gamma_4/t-\frac{i}2 k\llap/ (2\alpha-1)}
-\frac{i}2 k\llap/ \right] \biggr\}_{\rm lin.} u(k),
\label{Pfres}\end{eqnarray}
with $\alpha\equiv\sigma_1-\sigma_2$ and $\sigma_i\equiv\tau_i/t$.

Because $k_4=\frac{2\pi}{\beta}(n_k+\frac12)$, we find that the
integrand of Eq.~(\ref{Pfres}) is symmetric under
$\alpha\leftrightarrow 1-\alpha$,
since then $e^{-in\beta k_4\alpha} \leftrightarrow
(-1)^n e^{in\beta k_4\alpha}$. Therefore we can replace
$\int_0^1 d\sigma_1 \int_0^{\sigma_1}d\sigma_2 \to
\frac12\int_0^1 d\alpha$. The same substitution, when
applied to only part of the integrand, allows us
to write Eq.~(\ref{Pfres}) either as a straight or an alternating
sum over winding numbers $n$, which can be written in terms of
Jacobi $\theta$ functions, Eq.~(\ref{th234}), as
\begin{mathletters}
\label{ffinres}
\begin{eqnarray}
&&{\rm Str}\exp(-{\sf H}t/2)\nonumber\\
&=&\frac{\lambda^2}{4(2\pi t)^2}\int_0^1 d\alpha
e^{-\frac12 k^2 \alpha(1-\alpha)t}\bar u(-k)
\left[ \theta_3\Bigl(-\frac{\beta k_4 \alpha}{2\pi}
+\frac{\beta\gamma_4}{2\pi it} \Big| \frac{i\beta^2}{2\pi t}\Bigr)
e^{-i k\llap/(1-\alpha)} \right]_{\rm lin.} u(k) \\
&=&\frac{\lambda^2}{4(2\pi t)^2}\int_0^1 d\alpha
e^{-\frac12 k^2 \alpha(1-\alpha)t}\bar u(-k)
\left[ \theta_4\Bigl(-\frac{\beta k_4 \alpha}{2\pi}
-\frac{\beta\gamma_4}{2\pi it} \Big| \frac{i\beta^2}{2\pi t}\Bigr)
e^{-i k\llap/\alpha} \right]_{\rm lin.} u(k),
\end{eqnarray}
where because only the terms linear in $\gamma$ matrices are to be
kept we can treat them as if they were c-numbers.
Each of these comparatively compact expressions can easily
be shown to be equivalent to the standard result Eq.~(\ref{fres})
by means of the Jacobi transformation, Eq.~(\ref{Jit}). The two
representations of Eq.~(\ref{ffinres}) are thus related
to sums over integral and half-integral Matsubara frequencies,
respectively,
corresponding to the two possible ways of reading the standard expression
primarily as a sum over the internal bosonic or fermionic frequencies.
\end{mathletters}

\narrowtext
\section{Summary}

We have shown that the alternative method of evaluation of
Green's functions based on the standard formulae for
quantum mechanical path integrals can be adapted to
the case of thermal Green's functions involving fermions,
provided that a separation is made between purely bosonic
and purely fermionic contributions. This separation is
a natural one when the generating functional is formulated
in terms of supermatrices. The resulting quantum mechanical
path integrals are topologically constrained, the one
associated with fermionic contributions picking up a minus
sign for increasing the winding number by one.
The evaluation of Green's functions along these lines leads to
Feynman-parametrized results, the Feynman parameters being
related to the points of insertion of external lines,
but no loop-momentum integral is encountered. In place
of the usual sum over Matsubara frequencies, however,
a sum over winding numbers emerges. The latter has been
shown in Ref.~\cite{I} to be more directly amenable to
a high-temperature expansion. The generalization of the
results presented here for a simple four-dimensional model
of scalars with Yukawa coupling to fermions
to more complicated ones can performed with the techniques
developed in Ref.~\cite{G}.

\narrowtext

\acknowledgments
D.\ G.\ C.\ M.\ would like to acknowledge NSERC for financial support
and to thank the Institute for Theoretical Physics of the
Technical University of Vienna, where part of this
work was done, for hospitality and financial assistance.
R.\ and D.\ McKenzie had some helpful suggestions at the outset
of this investigation.

\end{document}